%% file: FastTSCF.tex
\documentclass[conference]{IEEEtran}

\usepackage{cite}
\usepackage{url}

\usepackage{graphicx}
\usepackage{xcolor}
\usepackage{tikz}
\usetikzlibrary{arrows,positioning,shapes.geometric,fit,spy,matrix,patterns, backgrounds}
\usepackage{pgfplots}
\usepgfplotslibrary{groupplots}
\usepackage{amssymb}
\usepackage{amsmath}
\usepackage{cases}
\usepackage{bm}
\usepackage{amsthm}
\usepackage{booktabs}

\usepackage{array}
\usepackage{adjustbox}
\usepackage{colortbl}
\usepackage{lipsum}
\usepackage[linesnumbered,algoruled,boxed,lined]{algorithm2e}
\usepackage[normalem]{ulem}

\usepackage{xpatch}
\xpatchcmd{\NCC@ignorepar}{%
\abovedisplayskip\abovedisplayshortskip}
{%
\abovedisplayskip\abovedisplayshortskip%
\belowdisplayskip\belowdisplayshortskip}
{}{}

\DeclareMathOperator*{\argmin}{arg~min}
\DeclareMathOperator*{\sgn}{sgn}
\DeclareMathOperator*{\erfc}{erfc}

\newcommand{\fixme}[2]{\ifx&#2&{\leavevmode\color{red}#1}\else{\leavevmode\color{red}FIXME\{}#1{\leavevmode\color{red}\}}\footnote{{\leavevmode\color{red}#2}}\PackageWarning{Fixme}{#1: #2}\fi}

\title{Fast Thresholded SC-Flip Decoding of Polar Codes}

\author{\IEEEauthorblockN{Furkan Ercan and Warren J. Gross}
\IEEEauthorblockA{Department of Electrical and Computer Engineering, McGill University, Montr\'eal, Qu\'ebec, Canada\\
Email: furkan.ercan@mail.mcgill.ca, warren.gross@mcgill.ca}}

\input{colors}

\begin{document}

\maketitle

\begin{abstract} 
SC-Flip (SCF) decoding algorithm shares the attention with the common polar code decoding approaches due to its low-complexity and improved error-correction performance. However, the inefficient criterion for locating the correct bit-flipping position in SCF decoding limits its improvements. Due to its improved bit-flipping criterion, Thresholded SCF (TSCF) decoding algorithm exhibits a superior error-correction performance and lower computational complexity than SCF decoding. However, the parameters of TSCF decoding depend on multiple channel and code parameters, and are obtained via Monte-Carlo simulations. Our main goal is to realize TSCF decoding as a practical polar decoder implementation. To this end, we first realize an approximated threshold value that is independent of the code parameters and precomputations. The proposed approximation has negligible error-correction performance degradation on the TSCF decoding. Then, we validate an alternative approach for forming a critical set that does not require precomputations, which also paves the way to the implementation of the Fast-TSCF decoder. Compared to the existing fast SCF implementations, the proposed Fast-TSCF decoder has $0.24$ to $0.41$ dB performance gain at frame error rate of $10^{-3}$, without any extra cost. Compared to the TSCF decoding, Fast-TSCF does not depend on precomputations and requires $87\%$ fewer decoding steps. Finally, implementation results in TSMC 65nm CMOS technology show that the Fast-TSCF decoder is $20\%$ and $82\%$ more area-efficient than the state-of-the-art fast SCF and fast SC-List decoder architectures, respectively.
\end{abstract}

\section{Introduction}

Polar codes, introduced by Ar{\i}kan, are a class of linear block codes that can provably achieve channel capacity of binary memoryless symmetric channels \cite{arikan09}, discrete and continuous memoryless channels \cite{sasoglu_polar09} under successive cancellation (SC) decoding, which have simple encoding/decoding properties. Due to such attractive properties of polar codes, they have been selected as the coding scheme for the control channel for enhanced mobile broadband (eMBB) use case within the $5^{\text{th}}$ generation wireless communication protocol (5G) \cite{38.212}. However, at practical code lengths, SC decoding yields mediocre error-correction performance. Besides, its sequential decoding nature is a limiting factor for throughput. To improve the error-correction performance of polar codes, enhanced decoding algorithms that use SC decoding as a base algorithm have been proposed. Among such algorithms, SC-List (SCL) decoding uses a list of SC decoders in parallel to keep track of up to $L$ best decoding paths throughout the decoding process \cite{TalList}. As a tradeoff for the improved error-correction performance, the list size $L$ adversely affects the computational complexity of SCL decoding \cite{ercan-allerton}. 

SC-Flip (SCF) decoding algorithm \cite{SCFlip14} uses additional SC decoding attempts in series in the case when an initial SC decoding fails due to a single channel-induced error. During the course of the initial SC decoding, a set of bit-flipping indices are calculated and stored based on a selection criterion. The average computational complexity of SCF is similar to that of SC at moderate-to-high signal-to-noise ratio ($E_b/N_0$) values and has improved error-correction performance comparable to SCL decoding with small list sizes. However, it was shown in \cite{SCF-WCNC18} that the bit-flipping criterion of SCF decoding is suboptimal, adversely impacting its average decoding complexity and error-correction performance. To address this issue, Thresholded SCF (TSCF) decoding was introduced in \cite{SCFlip_TCOM18} which uses a subset of indices that are most probable to hold an erroneous decision, called the \textit{critical set}, to reduce the computational effort on finding the correct bit-flipping index in the polar code. Moreover, a pre-computed and optimized threshold value serving as a criterion for finding the correct bit flipping position is shown to have improved error-correction performance compared to SCF decoding. Though, the main obstacles to the practical implementation of TSCF decoding are its lengthy, off-line precomputations that are required to establish a critical set and to find the optimum threshold value. Moreover, the current scheme of TSCF decoding has its optimum threshold value depends on code and channel parameters.

To tackle the problem of limited throughput of SC decoders, the identification of special bit-patterns that are found in polar codes are addressed in \cite{SSC2011,sarkis14}. Fast decoding of such special bit-patterns is then extended to SCF decoding in \cite{FastSCFlip_WCNC18,FastSCF-TCAS-I}. With the Fast-SCF decoding implementation in \cite{FastSCF-TCAS-I}, it was also shown that the special bit-patterns are also useful for limiting the search span of bit-flipping indices.

Our goal in this work is to realize TSCF decoding as a practical polar decoder implementation. To this end, we first observe how the code length, code rate and $E_b/N_0$ impacts the optimum value of the threshold, using the 5G polar codes. Based on the developed insight on the impact of the threshold value on the error-correction performance, we show that an approximation with negligible performance loss is possible. In return, the approximated threshold is a linear function of $E_b/N_0$, and is independent of code parameters and the precomputations. Then, by utilizing the theoretical performance bound of SC decoding as an evaluation metric, we show that an alternative non-empirical method for constructing a critical set can be replaced by the existing one. This adaptation does not only make the critical set of TSCF independent of associated lengthy precomputations but also paves the way towards the incorporation of fast decoding techniques within TSCF, to create the Fast-TSCF decoder. Finally, we implement the Fast-TSCF decoder with all the proposed simplifications and adaptations, using TSMC 65nm CMOS technology process. Compared to the state-of-the-art Fast-SCF decoder implementation, the proposed decoder has up to $0.29$ dB performance gain, is marginally faster, and has $20\%$ better area-efficiency. Compared to the baseline TSCF decoder, the proposed Fast-TSCF decoder requires an average of $8.2\times$ less decoding steps while exhibiting a similar error-correction performance, and does not require lengthy precomputations.

The rest of this work is organized as follows: Polar codes and decoding algorithms are detailed in Section~\ref{sec:bg}. Approximation on the threshold value and validation of the critical set construction for TSCF decoding are described in Section~\ref{sec:newTSCF}. Details on how to implement TSCF decoding for special nodes to establish Fast-TSCF is explained in Section~\ref{sec:FTSCF}. Comparative simulation and hardware synthesis results are depicted in Section~\ref{sec:res}, followed by concluding remarks in Section~\ref{sec:concl}.

\section{Preliminaries}\label{sec:bg}

\subsection{Polar Codes}\label{sec:bg:polar}
A polar code $PC(N,K)$ utilizes the phenomenon of channel polarization, that splits $N$ channels into $K$ reliable ones used to transmit the information bits, and $N-K$ unreliable ones, which are frozen to a known constant, usually to zero. The rate of a polar code is calculated as $R=K/N$. The set of frozen and non-frozen indices are represented with $\mathcal{A}^C$ and $\mathcal{A}$, respectively. 

The encoding of a polar code can be explained with a matrix multiplication, such that
\begin{equation}\label{eqn:enc}
\boldsymbol{x_{0:N-1}} = \boldsymbol{u_{0:N-1}}\boldsymbol{G}^{\otimes n}\text{,}
\end{equation}
where $\boldsymbol{x_{0:N-1}} = \{x_0,x_1,\ldots,x_{N-1}\}$ is the encoded vector, $\boldsymbol{u_{0:N-1}} = \{u_0,u_1,\ldots,u_{N-1}\}$ is the input vector, and the generator matrix $\boldsymbol{G}^{\otimes n}$ is the $n$-th Kronecker product ($\otimes$) of the binary polar code kernel $\boldsymbol{G} =  \left[\begin{smallmatrix} 1&0\\ 1&1 \end{smallmatrix} \right]$. In this context, $N = 2^n$, $n \in \mathbb Z^+$, and a polar code with length $N$ can be reinterpreted as a composition of two polar codes of length $N/2$. Throughout this work, we use the 5G polar code sequence, defined in \cite{38.212}.


\subsection{Successive Cancellation Decoding}\label{sec:bg:sc}

In SC decoding the bit estimation is performed in a sequential fashion, beginning from $\hat{u}_0$ towards $\hat{u}_{N-1}$. This is because the estimation of each bit depends on the channel observation $\boldsymbol{y}$ and previously estimated bits $\boldsymbol{\hat{u}_{0:i-1}}$, such that

\begin{equation}\label{eqn:scdecode}
\hat{u}_i=\left\{
  \begin{array}{@{}ll@{}}
    0, & \text{if } \text{Pr}[\boldsymbol{y},\boldsymbol{\hat{u}_{0:i\text{-}1}} | u_i = 0] \geq  \text{Pr}[\boldsymbol{y},\boldsymbol{\hat{u}_{0:i\text{-}1}} | u_i = 1]; \\
    0, & \text{if } i \in \mathcal{A}^C; \\
    1, & \text{otherwise.}
  \end{array}\right.
\end{equation}

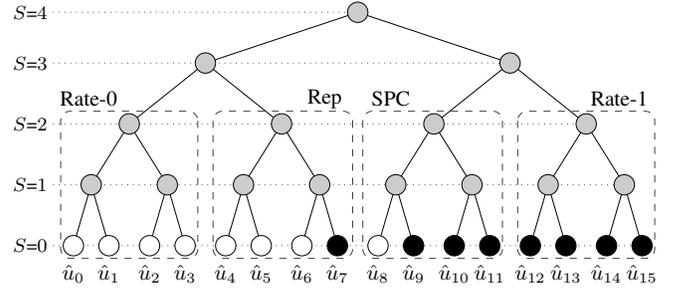
\begin{figure}
  \centering
  \scalebox{0.90}{
  \begin{tikzpicture}[scale=.75]
\usetikzlibrary{backgrounds}

\filldraw[fill=gray!40!white, draw=black] (+0.00,+0.00) circle [radius=.2];

\filldraw[fill=gray!40!white, draw=black] (-3.00,-1.00) circle [radius=.2];
\filldraw[fill=gray!40!white, draw=black] (+3.00,-1.00) circle [radius=.2];

\filldraw[fill=gray!40!white, draw=black] (-4.50,-2.20) circle [radius=.2];
\filldraw[fill=gray!40!white, draw=black] (-1.50,-2.20) circle [radius=.2];
\filldraw[fill=gray!40!white, draw=black] (+1.50,-2.20) circle [radius=.2];
\filldraw[fill=gray!40!white, draw=black] (+4.50,-2.20) circle [radius=.2];

\filldraw[fill=gray!40!white, draw=black] (-5.25,-3.40) circle [radius=.2];
\filldraw[fill=gray!40!white, draw=black] (-3.75,-3.40) circle [radius=.2];
\filldraw[fill=gray!40!white, draw=black] (-2.25,-3.40) circle [radius=.2];
\filldraw[fill=gray!40!white, draw=black] (-0.75,-3.40) circle [radius=.2];
\filldraw[fill=gray!40!white, draw=black] (+0.75,-3.40) circle [radius=.2];
\filldraw[fill=gray!40!white, draw=black] (+2.25,-3.40) circle [radius=.2];
\filldraw[fill=gray!40!white, draw=black] (+3.75,-3.40) circle [radius=.2];
\filldraw[fill=gray!40!white, draw=black] (+5.25,-3.40) circle [radius=.2];

\filldraw[fill=white!40!white, draw=black] (-5.60,-4.60) circle [radius=.2];
\filldraw[fill=white!40!white, draw=black] (-4.90,-4.60) circle [radius=.2];
\filldraw[fill=white!40!white, draw=black] (-4.10,-4.60) circle [radius=.2];
\filldraw[fill=white!40!white, draw=black] (-3.40,-4.60) circle [radius=.2];
\filldraw[fill=white!40!white, draw=black] (-2.60,-4.60) circle [radius=.2];
\filldraw[fill=white!40!white, draw=black] (-1.90,-4.60) circle [radius=.2];
\filldraw[fill=white!40!white, draw=black] (-1.10,-4.60) circle [radius=.2];
\filldraw[fill=black!40!black, draw=black] (-0.40,-4.60) circle [radius=.2];
\filldraw[fill=white!40!white, draw=black] (+0.40,-4.60) circle [radius=.2];
\filldraw[fill=black!40!black, draw=black] (+1.10,-4.60) circle [radius=.2];
\filldraw[fill=black!40!black, draw=black] (+1.90,-4.60) circle [radius=.2];
\filldraw[fill=black!40!black, draw=black] (+2.60,-4.60) circle [radius=.2];
\filldraw[fill=black!40!black, draw=black] (+3.40,-4.60) circle [radius=.2];
\filldraw[fill=black!40!black, draw=black] (+4.10,-4.60) circle [radius=.2];
\filldraw[fill=black!40!black, draw=black] (+4.90,-4.60) circle [radius=.2];
\filldraw[fill=black!40!black, draw=black] (+5.60,-4.60) circle [radius=.2];

\node [color=black] at (-5.60,-5.15) {\small $\hat{u}_0$};
\node [color=black] at (-4.90,-5.15) {\small $\hat{u}_1$};
\node [color=black] at (-4.10,-5.15) {\small $\hat{u}_2$};
\node [color=black] at (-3.40,-5.15) {\small $\hat{u}_3$};
\node [color=black] at (-2.60,-5.15) {\small $\hat{u}_4$};
\node [color=black] at (-1.90,-5.15) {\small $\hat{u}_5$};
\node [color=black] at (-1.10,-5.15) {\small $\hat{u}_6$};
\node [color=black] at (-0.40,-5.15) {\small $\hat{u}_7$};
\node [color=black] at (+0.40,-5.15) {\small $\hat{u}_8$};
\node [color=black] at (+1.10,-5.15) {\small $\hat{u}_9$};
\node [color=black] at (+1.90,-5.15) {\small $\hat{u}_{10}$};
\node [color=black] at (+2.60,-5.15) {\small $\hat{u}_{11}$};
\node [color=black] at (+3.40,-5.15) {\small $\hat{u}_{12}$};
\node [color=black] at (+4.10,-5.15) {\small $\hat{u}_{13}$};
\node [color=black] at (+4.90,-5.15) {\small $\hat{u}_{14}$};
\node [color=black] at (+5.60,-5.15) {\small $\hat{u}_{15}$};

\begin{scope}[on background layer]
\draw [-] (+0.00,+0.00) -- (-3.00,-1.00);
\draw [-] (+0.00,+0.00) -- (+3.00,-1.00);

\draw [-] (-3.00,-1.00) -- (-4.50,-2.20);
\draw [-] (-3.00,-1.00) -- (-1.50,-2.20);
\draw [-] (+3.00,-1.00) -- (+1.50,-2.20);
\draw [-] (+3.00,-1.00) -- (+4.50,-2.20);

\draw [-] (-4.50,-2.20) -- (-5.25,-3.40);
\draw [-] (-4.50,-2.20) -- (-3.75,-3.40);
\draw [-] (-1.50,-2.20) -- (-2.25,-3.40);
\draw [-] (-1.50,-2.20) -- (-0.75,-3.40);
\draw [-] (+1.50,-2.20) -- (+0.75,-3.40);
\draw [-] (+1.50,-2.20) -- (+2.25,-3.40);
\draw [-] (+4.50,-2.20) -- (+3.75,-3.40);
\draw [-] (+4.50,-2.20) -- (+5.25,-3.40);

\draw [-] (-5.25,-3.40) -- (-5.60,-4.60);
\draw [-] (-5.25,-3.40) -- (-4.90,-4.60);
\draw [-] (-3.75,-3.40) -- (-4.10,-4.60);
\draw [-] (-3.75,-3.40) -- (-3.40,-4.60);
\draw [-] (-2.25,-3.40) -- (-2.60,-4.60);
\draw [-] (-2.25,-3.40) -- (-1.90,-4.60);
\draw [-] (-0.75,-3.40) -- (-1.10,-4.60);
\draw [-] (-0.75,-3.40) -- (-0.40,-4.60);
\draw [-] (+0.75,-3.40) -- (+0.40,-4.60);
\draw [-] (+0.75,-3.40) -- (+1.10,-4.60);
\draw [-] (+2.25,-3.40) -- (+1.90,-4.60);
\draw [-] (+2.25,-3.40) -- (+2.60,-4.60);
\draw [-] (+3.75,-3.40) -- (+3.40,-4.60);
\draw [-] (+3.75,-3.40) -- (+4.10,-4.60);
\draw [-] (+5.25,-3.40) -- (+4.90,-4.60);
\draw [-] (+5.25,-3.40) -- (+5.60,-4.60);

\draw [-,dotted,color=white!30!black] (-6.00,+0.00) -- (+0.00,+0.00);
\draw [-,dotted,color=white!30!black] (-6.00,-1.00) -- (+3.00,-1.00);
\draw [-,dotted,color=white!30!black] (-6.00,-2.20) -- (+4.50,-2.20);
\draw [-,dotted,color=white!30!black] (-6.00,-3.40) -- (+5.25,-3.40);
\draw [-,dotted,color=white!30!black] (-6.00,-4.60) -- (+5.60,-4.60);

\end{scope}

\node [color=black] at (-6.45,+0.00) {\footnotesize $S$=$4$};
\node [color=black] at (-6.45,-1.00) {\footnotesize $S$=$3$};
\node [color=black] at (-6.45,-2.20) {\footnotesize $S$=$2$};
\node [color=black] at (-6.45,-3.40) {\footnotesize $S$=$1$};
\node [color=black] at (-6.45,-4.60) {\footnotesize $S$=$0$};

\draw [rounded corners, dashed, color=darkgray](+3.15,-1.95) rectangle (+5.85,-4.85); 
\draw [rounded corners, dashed, color=darkgray](+0.10,-1.95) rectangle (+2.85,-4.85); 
\draw [rounded corners, dashed, color=darkgray](-3.15,-1.95) rectangle (-5.85,-4.85); 
\draw [rounded corners, dashed, color=darkgray](-0.10,-1.95) rectangle (-2.85,-4.85); 

\node [color=black] at (-5.30,-1.70) {\small Rate-0};
\node [color=black] at (-0.65,-1.70) {\small Rep};
\node [color=black] at (+0.65,-1.70) {\small SPC};
\node [color=black] at (+5.15,-1.70) {\small Rate-1};

\end{tikzpicture}
}
  \caption{Successive cancellation decoding tree for $PC(16,8)$. Stages ($S$) for each level and the sub-codes with special frozen bit-patterns (Rate-0, Rate-1, Rep, SPC) are outlined for reference.}
  \label{fig:scdecode_n16}
\end{figure}

SC decoding of polar codes can be translated as a binary tree search starting from the root node located at stage $S = n$ and priority given to the left branch, as depicted in Fig.~\ref{fig:scdecode_n16}. The noisy channel observation, quantified in log-likelihood ratio (LLR) form, is located at the root node. The LLRs are propagated towards the leaf nodes using
\begin{align}
{\alpha}^l_i &= \sgn(\alpha^{v}_{i})\sgn(\alpha^{v}_{i+2^{S-1}}) \min(|\alpha^{v}_{i}|,|\alpha^{v}_{i+2^{S-1}}|) \text{,} \label{eqn:alphaleft}\\
{\alpha}^r_i &= \alpha^{v}_{i+2^{S-1}} + (1-2\beta^{l}_{i})\alpha^{v}_{i} \text{.} \label{eqn:alpharight}
\end{align}
where $\boldsymbol{\alpha^{v}}$ is the LLR of the parent node located at stage $S$, $\boldsymbol{\alpha^{l}}$ and $\boldsymbol{\alpha^{r}}$ are the LLRs of the left and right child nodes, respectively. The partial sums $\boldsymbol{\beta}$ observed from the left and right child nodes are passed to their parent nodes as
\begin{equation}\label{eqn:beta}
  \beta^{v}_i=\left\{
  \begin{array}{@{}ll@{}}
    \beta^{l}_{i} \oplus \beta^{r}_{i}, & \text{if}~ i \leq 2^{S-1} \\
    \beta^{r}_{i-2^{S-1}}, & \text{otherwise.}
  \end{array}\right.
\end{equation} 
where $\boldsymbol{\oplus}$ denotes bitwise XOR operation, and $0 \leq i < 2^S$. 

It was shown in \cite{SSC2011} and \cite{sarkis14} that dedicated fast decoding techniques at special sub-trees (\textit{i.e.} nodes) with unique frozen leaf node patterns improves the throughput of SC decoding tremendously. Within the 5G polar code sequence, identified special frozen leaf node patterns for polar codes are Rate-0 (where all indices are frozen) Rate-1 (where no indices are frozen), Rep (where only the rightmost index is non-frozen), single parity check (SPC) (where only the leftmost index is frozen) and two other unique nodes with patterns ($0011$),($0101$) for which $0$ and $1$ represent frozen and non-frozen indices, respectively \cite{FastSCF-TCAS-I,Thibaud_Survey}.

\subsection{SC-Flip Decoding}\label{sec:bg:scf}

When SC decoding fails, the bit-wise mismatches in the estimated codeword are either due to channel noise, or propagated errors induced by an earlier error in the sequential decoding process. With correcting the first channel-induced error, its associated propagated errors also disappear from the estimated codeword, and the error-correction performance is improved. 

To this end, SCF decoding uses an outer cyclic redundancy check (CRC) in the encoding and decoding of polar codes, to tell whether an initial SC decoding has failed. In case of a failed decoding, a set of non-frozen flipping indices, are sorted and stored based on their LLR magnitude information, in increasing order. This is followed by a new attempt in decoding. At each extra decoding, the flipping index with the next lowest LLR magnitude is flipped, in an attempt to correct a single channel-induced error. The additional decoding attempts continue until either the CRC passes, or a maximum number ($T_{\text{max}}$) is reached.

The main problem of the SCF decoding is that the LLR magnitude information at the non-frozen leaf indices is not sufficient for efficient identification of the first channel-induced error in the codeword. As such, different SCF-based algorithms have emerged to tackle this problem. In \cite{PSCF-ICC18,PSCF_IEEEAccess2019_Li}, partitioned SCF decoding was introduced that divides the codeword into several partitions, each of which is equipped with CRC and SCF is executed within the partitions separately. This approach was shown to improve the error-correction performance and to reduce the average number of iterations. In \cite{SCFlip17-conf,SCFlip17-jour}, dynamic SCF decoding introduced an improved metric that can efficiently identify the bit-flipping positions, and an algorithm to build the bit-flip list dynamically so that more than a single channel-induced error can be targeted. However, its metric calculation requires transcendental computations that are not suitable for practical implementations. In \cite{SCF-GLOBECOM17, ProgressiveSCF_IEEEAccess}, a subset of indices that are more likely to incur an error than others is created. Called \textit{critical set}, the indices are gathered using the first index of each Rate-1 node found in the decoding tree, and they substantially reduce the search span for the bit-flipping location. An independent but similar approach is taken in \cite{SCF-WCNC18} by considering a critical set based on the empirically-observed probabilities of channel-induced errors. 

Thresholded SCF (TSCF) decoding algorithm, introduced in \cite{SCFlip_TCOM18}, uses a critical set based on empirical studies. When the initial SC decoding fails, only the indices within the constructed critical set are evaluated for flipping. To reduce the search span further, a soft-value threshold $\Omega$ is applied, such that the indices that hold an LLR magnitude larger than $\Omega$ are not considered for bit-flipping. This approach was shown to improve the efficiency of the identification of first channel-induced error greatly, which in return improves the error-correction performance and reduces the average number of iterations. However, both the critical set and the threshold value are obtained using Monte-Carlo simulations which limits the practical implementation and flexibility of the TSCF algorithm.

\subsection{Fast-SCF Decoding}\label{sec:bg:fscf}

The identification of the special nodes and their fast SC decoding techniques  reported in \cite{SSC2011,sarkis14} were first embodied in SCF decoding in \cite{FastSCFlip_WCNC18}, and were improved in \cite{FastSCFlip_ICC19_Zhou, FastSCF-TCAS-I}. The main challenge on fast node decoding techniques is to identify and flip indices from the top of these special nodes and without explicitly traversing them. In \cite{FastSCF-TCAS-I}, it was shown that the fast decoding techniques specialized to support bit-flipping maintain a similar error-correction performance to SCF decoding. This idea was inspired after \cite{SCF-GLOBECOM17}, where a critical set is created using the first index of each Rate-1 node in the decoding tree. The LLR magnitude of the first index can also be found directly as the minimum LLR magnitude at a top index of a Rate-1 node, and flipping the top-node index holding the minimum LLR magnitude is sufficient to establish a decoding performance identical to SCF decoding. This approach is then extended to SPC nodes with flipping up to three indices, with which a similar error-correction performance is maintained. The computational complexity of the sorter unit is the main beneficiary of these simplifications, which often poses as a power and/or performance bottleneck in practical polar decoder applications \cite{alexios_sorter}.

\section{Towards Practical TSCF Decoding}\label{sec:newTSCF}

\subsection{Approximation of the LLR Threshold}\label{sec:FTSCF:omega}

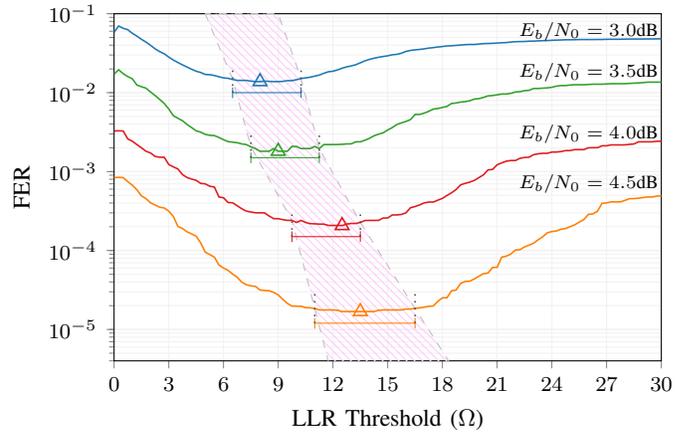
\begin{figure}
  \centering
  \input{figures/omega_vs_FER.tikz} 
  \vspace{-8mm}
  \caption{\label{fig:omega_vs_FER} An example on the impact of $\Omega$ on the FER of TSCF decoding, using $PC(256,128)$ and select $E_b/N_0$ values. CRC length is $C=16$ and $T_{\text{max}}=10$. Triangle markers at each curve represent $\Omega_{\text{opt}}$, and their associated range bars scale the range for $\Omega$ when a loss of up to $10\%$ from the optimum FER is tolerable.}
\end{figure}

In \cite{SCFlip_TCOM18}, an optimum value for the LLR threshold ($\Omega_{\text{opt}}$) is determined by simulations that are performed off-line and with respect to the code length, code rate and the $E_b/N_0$. 
Our goal is to be able to express $\Omega$ as a function of a simple parameter to avoid the associated off-line computations. Through Monte-Carlo simulations, we observed that the value of $\Omega$ is able to feature limited flexibility while maintaining a similar decoding performance.
Therefore, our idea is to tolerate a negligible performance loss, so that $\Omega_{\text{opt}}$ can be replaced with a range of $\Omega$ values. In other words, we claim that an acceptable $\Omega$ can be a range of values rather than a single point, and this flexibility paves the way for a regression study. 
To illustrate, Fig.~\ref{fig:omega_vs_FER} presents how the performance of TSCF decoding change with $\Omega$ at different $E_b/N_0$ values for $PC(256,128)$. The CRC polynomial is selected as $0\text{x}1021$ that has length $C=16$. The triangle markers on each curve show $\Omega_{\text{opt}}$ for each $E_b/N_0$ point, and the \textit{bars} under each curve represent the range the $\Omega$ can take, when up to $10\%$ loss \textit{from the optimum FER} is tolerated, \textit{e.g.} for $\text{FER}_{\text{opt}}=10^{-5}$, an error-correction performance of up to $\text{FER}=1.1 \times 10^{-5}$ is considered as tolerable. This tolerance reveals an area as depicted in Fig.~\ref{fig:omega_vs_FER}, within which a linear regression can be applied. 

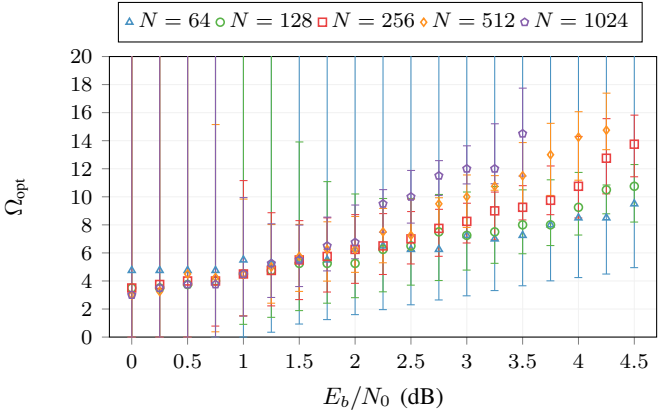
\begin{figure}
  \centering
  \input{figures/omega_vs_SNR.tikz} 
  \vspace{-7.5mm}
  \caption{\label{fig:omega_vs_SNR} $\Omega_{\text{opt}}$ as a function of $E_b/N_0$ and code length $N$, where code rate is fixed to $R = 0.5$ and CRC polynomial is $0x1021$. The bars associated with each $\Omega_{\text{opt}}$/$N$ point represent the flexibility of $\Omega$ when a performance loss of up to $10\%$ from the optimum FER value is tolerated.}
\end{figure}

\begin{figure}
  \centering
  \input{figures/omega_vs_R.tikz} 
  \vspace{-7.5mm}
  \caption{\label{fig:omega_vs_R} $\Omega_{\text{opt}}$ as a function of code rate $R$ and code length $N$, where $E_b/N_0 = 2.5$ dB and CRC polynomial is $0x1021$. The bars associated with each $\Omega_{\text{opt}}$/$N$ point represent the flexibility of $\Omega$ when a performance loss of up to $10\%$ from the optimum FER value is tolerated.}
\end{figure}
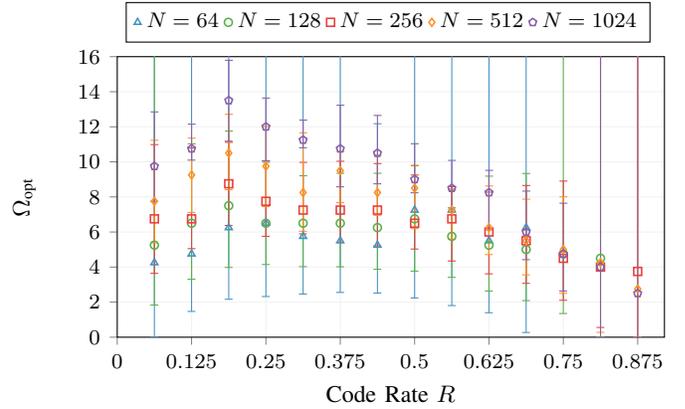

Fig.~\ref{fig:omega_vs_SNR} presents a broader perspective of the presented case, using the 5G polar code sequence. The markers show how the $\Omega_{\text{opt}}$ changes with the $N$ and $E_b/N_0$. The code rate is fixed to $R = 0.5$. The bars associated with each marker present the \textit{flexibility} for $\Omega$ when up to $10\%$ loss from the optimum FER is acceptable. 
In this sense, the bars provide an idea on by how much $\Omega_{\text{opt}}$ can be approximated without incurring a significant loss from FER performance.
According to Fig.~\ref{fig:omega_vs_SNR}, the flexibility of the $\Omega$ value is higher for smaller code lengths, and also at lower $E_b/N_0$ values. 
On the other hand, the lengths of the bars tend to shrink with the increasing $E_b/N_0$ for each code length, which means that the error-correction performance begins degrading with inconvenient $\Omega$ values. Lastly, the $\Omega_{\text{opt}}$ values for each considered length are clustered together and increase with $E_b/N_0$. 
Fig.~\ref{fig:omega_vs_R} depicts the $\Omega$ as a function of code rate $R$ for the same $N$ values, with $E_b/N_0 = 2.5$ dB. Similar to Fig.~\ref{fig:omega_vs_SNR}, bars extending from $\Omega_{\text{opt}}$ at each point represent the range for a tolerated loss of up to $10\%$ from the optimum FER. It can be seen that the $\Omega_{\text{opt}}$ values for each $N$ increase at moderate code rates and decrease at both lower and higher code rates. Fortunately, the ranges associated with each $\Omega$ are mostly wider at lower and higher code rates, which allows for a constant approximation.

Based on the findings derived from Fig.~\ref{fig:omega_vs_SNR} and Fig.~\ref{fig:omega_vs_R}, we propose an \textit{approximate threshold} $\Omega^{*}$ that is based on a linear regression model (\textit{i.e.} $f(x)=ax+b$). According to our comprehensive Monte-Carlo studies, our findings illustrated in Fig.~\ref{fig:omega_vs_SNR} and Fig.~\ref{fig:omega_vs_R}, and accounting for a hardware-friendly quantization, the linear regression for $\Omega^{*}$ is given as
\begin{equation}\label{eqn:omegaapprox}
\Omega^{*}(x) = 2(x + 3)
\end{equation}
where $x$ is the $E_b/N_0$ value in dB. 

\begin{figure}[t]
  \centering
  \input{figures/approxOmega_demo3.tikz}
  \vspace{-7.5mm}
  \caption{\label{fig:approxOmega_demo} FER comparison of TSCF decoding when $\Omega^{*}$ is used against the original case with $\Omega_{\text{opt}}$, using several different polar code lengths and code rates. $C=16$, $T_{\text{max}}=10$.}
\end{figure}
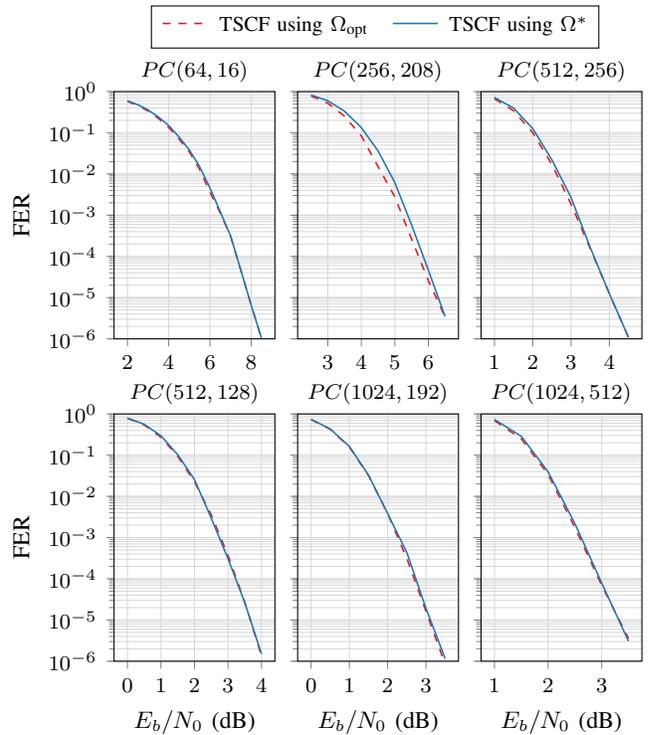

In order to validate our approximated approach, we select several polar codes of different code lengths and code rates, and compare their error-correction performance when using $\Omega_{\text{opt}}$ and when using $\Omega^{*}$. Fig.~\ref{fig:approxOmega_demo} compares the error-correction performance of TSCF algorithm when $\Omega^{*}$ is replaced with $\Omega_{\text{opt}}$ for six different polar codes. Four of the selected polar codes, located on the left of Fig.~\ref{fig:approxOmega_demo} are selected purposefully since their $\Omega_{\text{opt}}$ values are placed relatively far from the proposed hardware-friendly regression. It can be seen that the TSCF decoding with the approximation has FER performance trends very close to the original scheme. The worst-case performance loss due to $\Omega^{*}$ is about $0.02$ dB at FER$=10^{-3}$ with $PC(512,256)$, and about $0.1$ dB at FER$=10^{-4}$ with $PC(256,208)$. 

\subsection{Correlation of the Critical Sets}\label{sec:FTSCF:cs}

The critical set in \cite{SCF-WCNC18} is constructed empirically, and the critical set in \cite{SCF-GLOBECOM17} is constructed based on the code construction. We denote these critical sets based on their references, as $\mathcal{C}_{\text{\cite{SCF-WCNC18}}}$ and $\mathcal{C}_{\text{\cite{SCF-GLOBECOM17}}}$, respectively. Both approaches to critical set construction attempt to enclose the channel-induced error-prone indices; however their contents do not correlate perfectly. For example, for the $PC(16,8)$ code in Fig.~\ref{fig:scdecode_n16}, both $\mathcal{C}_{\text{\cite{SCF-WCNC18}}}$ and $\mathcal{C}_{\text{\cite{SCF-GLOBECOM17}}}$ contain the indices $\{u_{7}, u_{9}, u_{10}, u_{12}\}$, but $\mathcal{C}_{\text{\cite{SCF-WCNC18}}}$ also contains the indices $\{u_{11}, u_{13}\}$. The extra indices involved in $\mathcal{C}_{\text{\cite{SCF-WCNC18}}}$ are less likely to incur a channel-induced error in general; however, their inclusion (or exclusion) to the critical set alters the error-correction performance.

Our goal is to replace the empirically-obtained critical set of TSCF decoding with the systematic method presented in \cite{SCF-GLOBECOM17}, to remove the effort of off-line critical set construction, and also to enable implementation of Fast-TSCF decoding. To validate the correlation of the critical sets, we evaluate them using theoretical FER calculation. It was shown in \cite{wu2014construction} that a theoretical FER for SC decoding can be derived by using the error probability of each sub-channel (\textit{i.e.} leaf node index) of a polar code, such that
\begin{equation}\label{eqn:FERtheory}
\text{FER}_{\text{SC}} = 1- \Big[ \prod_{i \in \mathcal{A}}(1-\pi_i) \Big] 
\end{equation}
where $\pi_i$ denotes the probability of a channel-induced error at index $i$. 
For an AWGN channel with BPSK modulation, and with assuming all-zero codeword, the LLR at an index $i$ is represented as a Gaussian random variable with mean $\mu$ and variance $\sigma^2$, its associated $\pi_i$ can be approximated using the complementary error function \cite{Trifonov_polarcodedesign2012}:
\begin{equation}\label{eqn:pi}
\pi_i \approx \frac{1}{2} \erfc \Big(\frac{\sqrt{\mu_i}}{2}\Big)
\end{equation}

\begin{figure}[t]
  \centering
  \input{figures/Rate_vs_FERtheoretical2.tikz}
  \vspace{-2.5mm}
  \caption{\label{fig:Rate_vs_FERtheoretical} Theoretical FER used as a correlation and validation metric for the critical set construction approaches from \cite{SCF-WCNC18} and \cite{SCF-GLOBECOM17}, using (\ref{eqn:FERtheory}) and (\ref{eqn:FERcrit}). A rate and $E_b/N_0$ sweep (which increases in the outlined direction) is performed for 5G polar code of length $N=1024$, without an outer CRC.}
\end{figure}
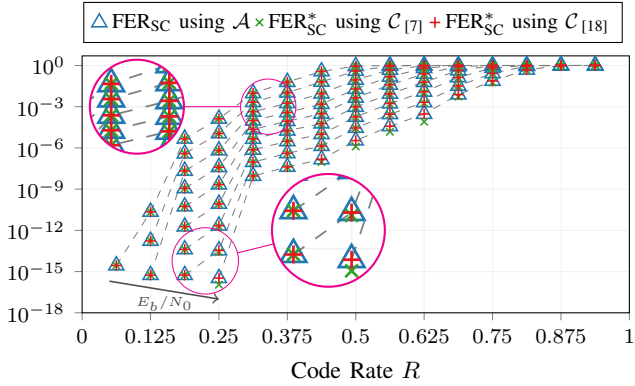

Equation (\ref{eqn:FERtheory}) can be elaborated for a theoretical FER bound for SCF decoding \cite{Fazeli_Viterbi}, however their approximations are known to be loose compared to their simulated counterparts (see Section IV-B of \cite{SCFlip17-jour} for more details). Consequently, we use the theoretical FER for SC decoding to demonstrate the correlation of the critical sets. Based on (\ref{eqn:FERtheory}), we introduce a \textit{hypothetical FER} created by replacing the non-frozen set with the critical set, such that
\begin{equation}\label{eqn:FERcrit}
\text{FER}^{*}_{\text{SC}} = 1- \Big[ \prod_{i \in \mathcal{C}}(1-\pi_i) \Big] 
\end{equation}
where $\mathcal{C}$ denotes the critical set. In this context, with a well-constructed critical set, $\text{FER}^{*}_{\text{SC}}$ should be almost identical to $\text{FER}_{\text{SC}}$. Fig.~\ref{fig:Rate_vs_FERtheoretical} presents the theoretical FER comparison for (\ref{eqn:FERtheory}) with (\ref{eqn:FERcrit}) using $\mathcal{C}_{\text{\cite{SCF-WCNC18}}}$, and (\ref{eqn:FERcrit}) using $\mathcal{C}_{\text{\cite{SCF-GLOBECOM17}}}$. 
The $\mu_i$ in (\ref{eqn:pi}) is computed using Gaussian approximation \cite{Trifonov_polarcodedesign2012} for the 5G polar code of length $1024$ and several $E_b/N_0$ values. 
According to Fig.~\ref{fig:Rate_vs_FERtheoretical}, while both critical set construction methods correlate well at low and moderate $E_b/N_0$ values, the critical set $\mathcal{C}_{\text{\cite{SCF-WCNC18}}}$ starts to lose  its precision at high $E_b/N_0$ values. This is because the limited amount of precomputations become insufficient to identify erroneous indices at higher $E_b/N_0$ values. On the other hand, the critical set $\mathcal{C}_{\text{\cite{SCF-GLOBECOM17}}}$ is able to maintain a well-approximated trend line and thus is more favorable for the implementation of critical set-based implementations such as TSCF decoding.

\section{Fast-TSCF Decoding}\label{sec:FTSCF}

The demonstrated correlation in Section~\ref{sec:FTSCF:cs} allows for the TSCF algorithm to be implemented using the special nodes in Fast-SCF decoding, which we name as the \textit{Fast-TSCF decoding} algorithm. Hence, we briefly review the evaluation of special nodes under Fast-TSCF decoding. In the following, let $\alpha^{S}_{0:N_v-1}$ denote the root LLR vector of the special node located at tree stage $S$ and has length $N_v$, and let $\eta$ denote the top-node bit-flipping index. 

At Rate-1 nodes, only one top-node index that holds the minimum LLR magnitude is considered for bit-flipping, if it is smaller than or equal to threshold $\Omega$, such that
\begin{equation}\label{eqn:node:r1}
\eta_{\text{Rate-1}} =\left\{
  \begin{array}{@{}ll@{}}
    \argmin |\alpha^{S}_{0:N_v-1}|, & \text{if}~ \min |\alpha^{S}_{0:N_v-1}| \leq \Omega \\
    \varnothing, & \text{otherwise.}
  \end{array}\right.
\end{equation}

At Rep nodes, the LLR magnitude of the only non-frozen leaf node index is directly obtained by summing all its top-node LLRs. If the corresponding LLR magnitude is smaller than the threshold, then the entire Rep node is evaluated for bit-flipping:
\begin{equation}\label{eqn:node:rep}
\eta_{\text{Rep}} =\left\{
  \begin{array}{@{}ll@{}}
    \forall i, & \text{if}~ \bigg{|} \sum_{\forall i \in N_v} \alpha^{S}_{i}\bigg{|} \leq \Omega \\
    \varnothing, & \text{otherwise.}
  \end{array}\right.
\end{equation}

SPC nodes can be considered as a composition of a single frozen index, followed by Rate-1 nodes of sizes ranging from $1$ to $N_v/2$. According to the critical set construction based on Rate-1 nodes, SPC nodes could contain more than one critical index, based on its size. The bit-flipping criterion of Rate-1 nodes in (\ref{eqn:node:r1}) cannot be applied within SPC nodes without a tree-traversal. Therefore, our approach is based on the simplified bit-flipping criteria in SPC nodes that were detailed in \cite{FastSCF-TCAS-I}: depending on the state of their parity ($p$), two subsets of top-node indices that hold the first, second or third minimum LLR magnitudes are evaluated for bit-flipping. In our case, if the sum of the LLR magnitudes of these indices is less than the threshold, they are considered for bit-flipping. Accordingly, if $i_\text{min,j}$ denotes the index with $j^{\text{th}}$ minimum LLR magnitude, our criteria for the aforementioned two bit-flipping subsets within SPC node are detailed as follows:
\begin{equation}\label{eqn:node:spc1}
\eta_{\text{1,SPC}} =\left\{
  \begin{array}{@{}lll@{}}
    i_\text{min,2}, & \text{if}~ p=1~\text{\&}~|\alpha^{S}_{i_\text{min,2}}| \leq \Omega \\
    \{i_\text{min,1},i_\text{min,2}\}, & \text{if}~ p=0~\text{\&}~|\alpha^{S}_{i_\text{min,1}}|+|\alpha^{S}_{i_\text{min,2}}| \leq \Omega \\
    \varnothing, & \text{otherwise.}
  \end{array}\right.
\end{equation}
\begin{equation}\label{eqn:node:spc2}
\eta_{\text{2,SPC}} =\left\{
  \begin{array}{@{}lll@{}}
    i_\text{min,3}, & \text{if}~ p=1~\text{\&}~|\alpha^{S}_{i_\text{min,3}}| \leq \Omega \\
    \{i_\text{min,1},i_\text{min,3}\}, & \text{if}~ p=0~\text{\&}~|\alpha^{S}_{i_\text{min,1}}|+|\alpha^{S}_{i_\text{min,3}}| \leq \Omega \\
    \varnothing, & \text{otherwise.}
  \end{array}\right.
\end{equation}

\section{Results}\label{sec:res}

\subsection{Error-Correction Performance}\label{sec:res:fer}

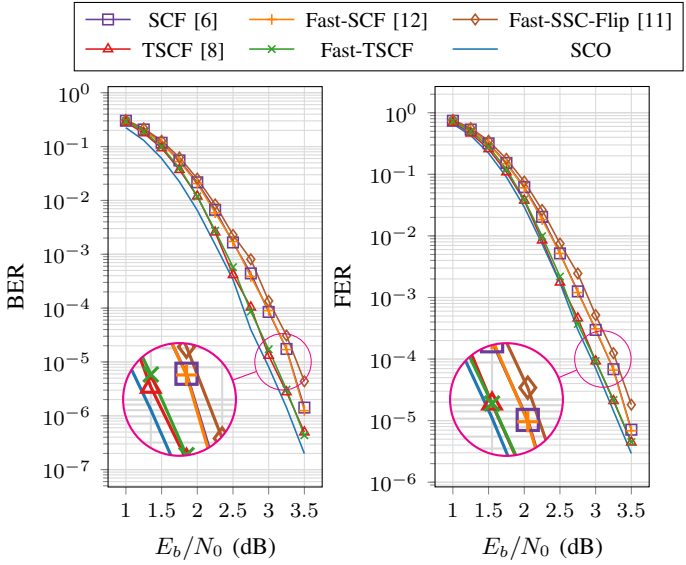
\begin{figure}
  \centering
  \input{figures/results_perf.tikz} 
  \vspace{-5mm}
  \caption{\label{fig:results_perf} Error-correction performance comparison for the proposed Fast-TSCF decoding against other SCF-based decoding algorithms.}
\end{figure}

The error-correction performance of the proposed Fast-TSCF decoding that uses the approximated threshold and the fast node decoding techniques is compared against TSCF \cite{SCFlip_TCOM18}, SCF \cite{SCFlip14}, Fast-SCF \cite{FastSCF-TCAS-I} and Fast-SSC-Flip \cite{FastSCFlip_WCNC18} in Fig.~\ref{fig:results_perf}, using $PC(1024,512)$ from the 5G polar code sequence, $C=16$, and $T_{\text{max}}=10$. A genie-aided decoder that always corrects the first channel error, called SC-Oracle (SCO) decoder, is also depicted to represent the lower bound for all SCF decoding algorithms. According to Fig.~\ref{fig:results_perf}, Fast-TSCF exhibits similar performance to that of TSCF, and its performance gain compared to SCF/Fast-SCF is $0.24$ dB at FER$=10^{-3}$ and $0.20$ dB at FER$=10^{-4}$. Compared to the Fast-SSC-Flip algorithm, the proposed decoder has a performance gain of $0.41$ dB at FER$=10^{-3}$ and $0.29$ dB at FER$=10^{-4}$. Finally, it is worth to mention that TSCF and Fast-TSCF decoders exhibit the closest error-correction performance to the SC-Oracle.

\subsection{Computational Complexity}\label{sec:res:iter}

\begin{figure}
  \centering
  \input{figures/results_iter.tikz} 
  \vspace{-7mm}
  \caption{\label{fig:results_iter} Comparison of average number of decoding steps for the proposed Fast-TSCF decoding against other SCF-based decoding algorithms.}
\end{figure}
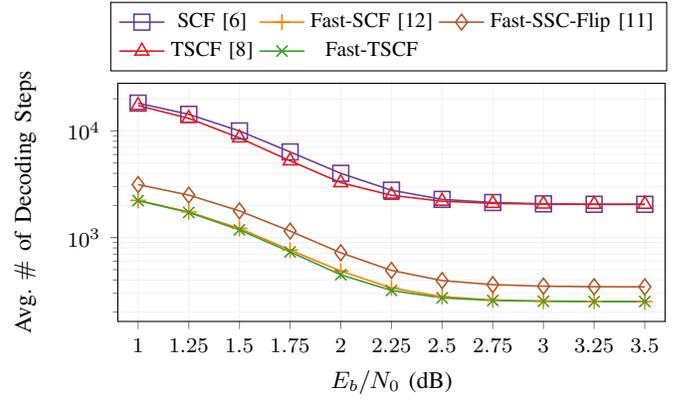

The average computational complexity for each SCF-based decoding is represented by the average number of decoding steps, which is calculated by multiplying the average number of iterations with the number of decoding steps performed to complete one full iteration. Fig.~\ref{fig:results_iter} depicts the comparison of the average number of decoding steps for the proposed Fast-TSCF decoder against the aforementioned SCF-based decoders, following the same setup to create Fig.~\ref{fig:results_perf}. In this regard, the number of decoding steps for Fast-SCF, Fast-SSC-Flip, and Fast-TSCF is calculated by following their fast node decoding techniques. For the SCF and TSCF decoding, $2N-2$ steps are considered for a full decoding iteration, following the definition in \cite{arikan09}. It can be seen from Fig.~\ref{fig:results_iter} that the proposed Fast-TSCF decoding exhibits the lowest average number of iterations, closely followed by Fast-SCF. Moreover, Fast-TSCF decoding requires $32\%$, $88\%$ and $87\%$ fewer decoding steps on average than Fast-SSC-Flip, SCF and TSCF, respectively. 

\subsection{ASIC Synthesis Results}\label{sec:res:asic}

The proposed Fast-TSCF decoder has been implemented in VHDL, validated with test benches and synthesized using Cadence Genus RTL compiler with the TSMC 65nm CMOS technology process. The hardware architecture for the Fast-SCF decoder that was proposed recently in \cite{FastSCF-TCAS-I} has been used as the baseline implementation for our work, with the following changes: Since the TSCF algorithm stores the flipping indices based on their regular order or appearance in the codeword and not based on their LLR magnitudes, the insertion sorter of Fast-SCF decoding is not required in the proposed architecture. The channel estimation is introduced as an input to the architecture, to be translated to the approximated $\Omega$ function, as described in Section~\ref{sec:newTSCF}. The quantization parameters for the channel and internal LLRs and parallelization factors for read/write operations from/to memories are kept the same as Fast-SCF for a fair comparison scheme.

Table~\ref{tab:results} presents the synthesis results for the proposed Fast-TSCF decoder. A comparison scheme is created against the state-of-the-art Fast-SCF decoder from \cite{FastSCF-TCAS-I}, and the fast SC-List decoder (Fast-SSCL) from \cite{fastSSCL-TSP} with a list size of $L=2$ since it is known that their error-correction performances are similar \cite{SCFlip_TCOM18}. Compared to the Fast-SCF decoder, the simplifications in the proposed architecture results in improved throughput and improved area efficiency. Compared to the Fast-SCF decoder, the proposed implementation is $5.5\%$ faster and requires $12.5\%$ less area, leading to $20\%$ more area-efficiency. Compared to the Fast-SSCL decoder, the proposed implementation has $14\%$ less average throughput but has $53\%$ less area, which leads to an overall area efficiency of $82\%$ more than that of Fast-SSCL.

\begin{table}[t]
\centering
\caption{TSMC 65 nm CMOS synthesis results comparison for Fast-TSCF decoding against state-of-the-art, using $PC(1024,512)$.}
\label{tab:results}
\resizebox*{1.00\columnwidth}{!}{
\begin{tabular}{@{}llll@{}}
\toprule
                     & Fast-TSCF & Fast-SCF \cite{FastSCF-TCAS-I} & Fast-SSCL\textsuperscript{(b)} \cite{fastSSCL-TSP} \\
                       \cmidrule(l){2-2} \cmidrule(l){3-3} \cmidrule(l){4-4}
Technology (nm)      & 65  & 65  & 65 \\
Supply(V)            & 1.0 & 1.0 & N/A \\
Frequency (MHz)      & 480 & 455 & 885 \\
Avg. Coded T/P (Mbps)	 & 1595\textsuperscript{(a)} & 1511\textsuperscript{(a)} & 1861\\
Area (mm$^2$)        & 0.49 & 0.56 & 1.05 \\
Area Efficiency (Gbps/mm$^2$) & 3.25 & 2.71 & 1.78 \\
\bottomrule
\multicolumn{3}{l}{\textsuperscript{(a)} Average value at target FER$=10^{-4}$.}\\
\multicolumn{3}{l}{\textsuperscript{(b)} List size is $L=2$.}\\
\end{tabular}}
\end{table}

\section{Conclusion}\label{sec:concl}

In this work, we proposed Fast-TSCF decoding that can compute the required critical set and the threshold value on-the-fly and incorporates fast decoding techniques. Specifically, we replaced the optimized thresholding of TSCF decoding which depends on code and channel parameters and requires lengthy precomputations, with an approximate threshold that is a function of a single channel parameter and does not require off-line computations. Then, by using the theoretical FER bound for SC decoding as an evaluation metric for validating critical sets, we showed that an alternative critical set can be employed by TSCF decoding that does not require precomputations. Finally, we introduced how to incorporate special nodes into TSCF decoding, which has led to the implementation of the Fast-TSCF decoder. Compared to the state-of-the-art Fast-SCF decoder, the proposed Fast-TSCF decoder has $0.24$ dB performance improvement at FER$=10^{-3}$, and synthesis results using TSMC 65nm CMOS technology process show that the proposed implementation exhibits $20\%$ more area-efficiency and improved throughput. Compared to the state-of-the-art fast SCL decoder, Fast-TSCF decoder is $82\%$ more area-efficient. Finally, compared to the baseline TSCF decoding, Fast-TSCF exhibits a similar error-correction performance but requires $88\%$ fewer decoding steps and has no pre-computational dependencies.

\section*{Acknowledgements}

The authors would like to thank Thibaud Tonnellier and Syed Mohsin Abbas for their helpful discussions and reviews.

\bibliographystyle{IEEEtran}

\end{document}

%% file: colors.tex
\definecolor{ocre}{RGB}{0, 157, 224} 
\definecolor{Paired-2}{RGB}{166,206,227}
\definecolor{Paired-1}{RGB}{31,120,180}
\definecolor{Paired-4}{RGB}{178,223,138}
\definecolor{Paired-3}{RGB}{51,160,44}
\definecolor{Paired-6}{RGB}{251,154,153}
\definecolor{Paired-5}{RGB}{227,26,28}
\definecolor{Paired-8}{RGB}{253,191,111}
\definecolor{Paired-7}{RGB}{255,127,0}
\definecolor{Paired-10}{RGB}{202,178,214}
\definecolor{Paired-9}{RGB}{106,61,154}
\definecolor{Paired-12}{RGB}{255,255,153}
\definecolor{Paired-11}{RGB}{177,89,40}
\definecolor{Accent-1}{RGB}{127,201,127}
\definecolor{Accent-2}{RGB}{190,174,212}
\definecolor{Accent-3}{RGB}{253,192,134}
\definecolor{Accent-4}{RGB}{255,255,153}
\definecolor{Accent-5}{RGB}{56,108,176}
\definecolor{Accent-6}{RGB}{240,2,127}
\definecolor{Accent-7}{RGB}{191,91,23}
\definecolor{Accent-8}{RGB}{102,102,102}
\definecolor{Spectral-1}{RGB}{158,1,66}
\definecolor{Spectral-2}{RGB}{213,62,79}
\definecolor{Spectral-3}{RGB}{244,109,67}
\definecolor{Spectral-4}{RGB}{253,174,97}
\definecolor{Spectral-5}{RGB}{254,224,139}
\definecolor{Spectral-6}{RGB}{255,255,191}
\definecolor{Spectral-7}{RGB}{230,245,152}
\definecolor{Spectral-8}{RGB}{171,221,164}
\definecolor{Spectral-9}{RGB}{102,194,165}
\definecolor{Spectral-10}{RGB}{50,136,189}
\definecolor{Spectral-11}{RGB}{94,79,162}
\definecolor{Set1-1}{RGB}{228,26,28}
\definecolor{Set1-2}{RGB}{55,126,184}
\definecolor{Set1-3}{RGB}{77,175,74}
\definecolor{Set1-4}{RGB}{152,78,163}
\definecolor{Set1-5}{RGB}{255,127,0}
\definecolor{Set1-6}{RGB}{255,255,51}
\definecolor{Set1-7}{RGB}{166,86,40}
\definecolor{Set1-8}{RGB}{247,129,191}
\definecolor{Set1-9}{RGB}{153,153,153}
\definecolor{Set2-1}{RGB}{102,194,165}
\definecolor{Set2-2}{RGB}{252,141,98}
\definecolor{Set2-3}{RGB}{141,160,203}
\definecolor{Set2-4}{RGB}{231,138,195}
\definecolor{Set2-5}{RGB}{166,216,84}
\definecolor{Set2-6}{RGB}{255,217,47}
\definecolor{Set2-7}{RGB}{229,196,148}
\definecolor{Set2-8}{RGB}{179,179,179}
\definecolor{Dark2-1}{RGB}{27,158,119}
\definecolor{Dark2-2}{RGB}{217,95,2}
\definecolor{Dark2-3}{RGB}{117,112,179}
\definecolor{Dark2-4}{RGB}{231,41,138}
\definecolor{Dark2-5}{RGB}{102,166,30}
\definecolor{Dark2-6}{RGB}{230,171,2}
\definecolor{Dark2-7}{RGB}{166,118,29}
\definecolor{Dark2-8}{RGB}{102,102,102}
\definecolor{Reds-1}{RGB}{255,245,240}
\definecolor{Reds-2}{RGB}{254,224,210}
\definecolor{Reds-3}{RGB}{252,187,161}
\definecolor{Reds-4}{RGB}{252,146,114}
\definecolor{Reds-5}{RGB}{251,106,74}
\definecolor{Reds-6}{RGB}{239,59,44}
\definecolor{Reds-7}{RGB}{203,24,29}
\definecolor{Reds-8}{RGB}{165,15,21}
\definecolor{Reds-9}{RGB}{103,0,13}
\definecolor{Greens-1}{RGB}{247,252,245}
\definecolor{Greens-2}{RGB}{229,245,224}
\definecolor{Greens-3}{RGB}{199,233,192}
\definecolor{Greens-4}{RGB}{161,217,155}
\definecolor{Greens-5}{RGB}{116,196,118}
\definecolor{Greens-6}{RGB}{65,171,93}
\definecolor{Greens-7}{RGB}{35,139,69}
\definecolor{Greens-8}{RGB}{0,109,44}
\definecolor{Greens-9}{RGB}{0,68,27}
\definecolor{Blues-1}{RGB}{247,251,255}
\definecolor{Blues-2}{RGB}{222,235,247}
\definecolor{Blues-3}{RGB}{198,219,239}
\definecolor{Blues-4}{RGB}{158,202,225}
\definecolor{Blues-5}{RGB}{107,174,214}
\definecolor{Blues-6}{RGB}{66,146,198}
\definecolor{Blues-7}{RGB}{33,113,181}
\definecolor{Blues-8}{RGB}{8,81,156}
\definecolor{Blues-9}{RGB}{8,48,107}
\definecolor{awesome-skyblue}{HTML}{0395DE}
\definecolor{bg}{HTML}{282828}

\definecolor{rosso}{RGB}{220,57,18}
\definecolor{giallo}{RGB}{255,153,0}
\definecolor{blu}{RGB}{102,140,217}
\definecolor{verde}{RGB}{16,150,24}
\definecolor{viola}{RGB}{153,0,153}

\definecolor{bleuUni}{RGB}{0, 157, 224}
\definecolor{marronUni}{RGB}{68, 58, 49}

%% file: figures/omega_vs_FER.tikz
\begin{tikzpicture}

    \begin{axis}[
            footnotesize, width=\columnwidth, height=0.70*\columnwidth,    
            xmin=0, xmax=30, xtick={0,3, ..., 30},
            ymin=4e-06,  ymax=1e-01,
            ymode=log,
            xlabel=LLR Threshold ($\Omega$), 
            ylabel=FER,  
            grid=both, grid style={gray!10},
            tick align=outside, tickpos=left, 
            legend columns =3,
            legend style={at={(0.495,1.05)},anchor=south}],
        ]

\draw [dashed, gray!50,pattern=north west lines, pattern color = magenta!20!white]  (axis cs:5.00,1.00e-1) -- (axis cs:6.50,1.47e-02) -- (axis cs:7.5,0.00205973) -- 
                 (axis cs:9.75,2.40955e-04) -- (axis cs:11,1.8455e-05) -- (axis cs:13,3e-7) -- 
                 (axis cs:20,1e-6) -- (axis cs:16.5,1.8455e-05) -- (axis cs:13.5,2.26644e-04) -- 
                 (axis cs:11.25,0.0019581) -- (axis cs:10.25,0.0151) -- (axis cs:9,1e-1) -- cycle;        
        
   	\addplot[Paired-1, semithick]
        table [x=OMEGA,y=FER,] {figures/data3/n256_k128_c16_t10_ebno3.0};
        
        \addplot[Paired-3, semithick]
        table [x=OMEGA,y=FER,] {figures/data3/n256_k128_c16_t10_ebno3.5};
        
        \addplot[Paired-5, semithick]
        table [x=OMEGA,y=FER,] {figures/data3/n256_k128_c16_t10_ebno4.0};
        
        \addplot[Paired-7, semithick]
        table [x=OMEGA,y=FER,] {figures/data3/n256_k128_c16_t10_ebno4.5};

        
        
        

        
        
        

        
        
%

        \addplot[Paired-1, semithick, mark=triangle, mark size=3,]
        table [x=OMEGA,y=FER,] {figures/data3/n256_k128_c16_t10_ebno3.0-optimumSolo};
        
        \addplot[Paired-3, semithick, mark=triangle, mark size=3,]
        table [x=OMEGA,y=FER,] {figures/data3/n256_k128_c16_t10_ebno3.5-optimumSolo};
        
        \addplot[Paired-5, semithick, mark=triangle, mark size=3,]
        table [x=OMEGA,y=FER,] {figures/data3/n256_k128_c16_t10_ebno4.0-optimumSolo};
        
        \addplot[Paired-7, semithick, mark=triangle, mark size=3,]
        table [x=OMEGA,y=FER,] {figures/data3/n256_k128_c16_t10_ebno4.5-optimumSolo};

        \addplot[Paired-1, semithick][error bars/.cd, x dir=both,x explicit] 
        table [x=OMEGA,y=FER, x error plus expr=\thisrow{LOW}-\thisrow{OMEGA},x error minus expr=\thisrow{OMEGA}-\thisrow{HIGH},] {figures/data3/n256_k128_c16_t10_ebno3.0-rangeBar};
        
        \addplot[Paired-3, semithick][error bars/.cd, x dir=both,x explicit] 
        table [x=OMEGA,y=FER, x error plus expr=\thisrow{LOW}-\thisrow{OMEGA},x error minus expr=\thisrow{OMEGA}-\thisrow{HIGH},] {figures/data3/n256_k128_c16_t10_ebno3.5-rangeBar};
        
        \addplot[Paired-5, semithick][error bars/.cd, x dir=both,x explicit] 
        table [x=OMEGA,y=FER, x error plus expr=\thisrow{LOW}-\thisrow{OMEGA},x error minus expr=\thisrow{OMEGA}-\thisrow{HIGH},] {figures/data3/n256_k128_c16_t10_ebno4.0-rangeBar};
        
        \addplot[Paired-7, semithick][error bars/.cd, x dir=both,x explicit] 
        table [x=OMEGA,y=FER, x error plus expr=\thisrow{LOW}-\thisrow{OMEGA},x error minus expr=\thisrow{OMEGA}-\thisrow{HIGH},] {figures/data3/n256_k128_c16_t10_ebno4.5-rangeBar};


\draw [-,dotted,color=black!70!white,semithick](axis cs:6.50,1.0e-02)--(axis cs:6.50,2.0e-02);
\draw [-,dotted,color=black!70!white,semithick](axis cs:10.25,1.0e-02)--(axis cs:10.25,2.0e-02);

\draw [-,dotted,color=black!70!white,semithick](axis cs:7.50,1.5e-03)--(axis cs:7.50,3.0e-03);
\draw [-,dotted,color=black!70!white,semithick](axis cs:11.25,1.5e-03)--(axis cs:11.25,3.0e-03);

\draw [-,dotted,color=black!70!white,semithick](axis cs:9.75,1.5e-04)--(axis cs:9.75,3.0e-04);
\draw [-,dotted,color=black!70!white,semithick](axis cs:13.5,1.5e-04)--(axis cs:13.5,3.0e-04);

\draw [-,dotted,color=black!70!white,semithick](axis cs:11,1.2e-05)--(axis cs:11,3.0e-05);
\draw [-,dotted,color=black!70!white,semithick](axis cs:16.5,1.2e-05)--(axis cs:16.5,3.0e-05);

\node [color=black] at (axis cs:26,6.0e-02) {\scriptsize $E_b/N_0=3.0$dB};
\node [color=black] at (axis cs:26,1.8e-02) {\scriptsize $E_b/N_0=3.5$dB};
\node [color=black] at (axis cs:26,3.0e-03) {\scriptsize $E_b/N_0=4.0$dB};
\node [color=black] at (axis cs:26,7.0e-04) {\scriptsize $E_b/N_0=4.5$dB};




    \end{axis}
\end{tikzpicture} 

%% file: figures/omega_vs_SNR.tikz
\begin{tikzpicture}
    \begin{axis}[
            footnotesize, width=\columnwidth, height=0.60\columnwidth,    
            xmin=-0.2, xmax=4.7, xtick={0,0.5, ..., 4.5},
            ymin=0,  ymax=20,
            xlabel=$E_b/N_0$ (dB),
            ylabel=$\Omega_{\text{opt}}$,  
            grid=both, grid style={gray!10},
            tick align=outside, tickpos=left, 
            legend columns =5,
            legend style={at={(0.495,1.05)},anchor=south}],
        ]
   	\addplot[only marks, Paired-1!70!Paired-2, mark=triangle, semithick, mark size=1.5][error bars/.cd, y dir=both,y explicit] 
        table [x=SNR,y=OPT, y error plus expr=\thisrow{LOW}-\thisrow{OPT},y error minus expr=\thisrow{OPT}-\thisrow{HIGH},] {figures/data/n64_k32_100in1000};

   	\addplot[only marks, Paired-3!70!Paired-4, mark=o, semithick, mark size=1.5][error bars/.cd, y dir=both,y explicit] 
        table [x=SNR,y=OPT, y error plus expr=\thisrow{LOW}-\thisrow{OPT},y error minus expr=\thisrow{OPT}-\thisrow{HIGH},] {figures/data/n128_k64_100in1000};

   	\addplot[only marks, Paired-5!70!Paired-6, mark=square, semithick, mark size=1.5][error bars/.cd, y dir=both,y explicit] 
        table [x=SNR,y=OPT, y error plus expr=\thisrow{LOW}-\thisrow{OPT},y error minus expr=\thisrow{OPT}-\thisrow{HIGH},] {figures/data/n256_k128_100in1000};

   	\addplot[only marks, Paired-7!70!Paired-8, mark=diamond, semithick, mark size=1.5][error bars/.cd, y dir=both,y explicit] 
        table [x=SNR,y=OPT, y error plus expr=\thisrow{LOW}-\thisrow{OPT},y error minus expr=\thisrow{OPT}-\thisrow{HIGH},] {figures/data/n512_k256_100in1000};
        
   	\addplot[only marks, Paired-9!70!Paired-10, mark=pentagon, semithick, mark size=1.5][error bars/.cd, y dir=both,y explicit] 
        table [x=SNR,y=OPT, y error plus expr=\thisrow{LOW}-\thisrow{OPT},y error minus expr=\thisrow{OPT}-\thisrow{HIGH},] {figures/data/n1024_k512_100in1000};

        \legend{$N=64$,$N=128$,$N=256$,$N=512$,$N=1024$}

    \end{axis}
\end{tikzpicture} 

%% file: figures/omega_vs_R.tikz
\begin{tikzpicture}
    \begin{axis}[
            footnotesize, width=\columnwidth, height=0.60\columnwidth,    
            xmin=0, xmax=0.92, xtick={0,0.125, ..., 0.875}, 
            xticklabel style={/pgf/number format/.cd,fixed,precision=3},
            ymin=0,  ymax=16,
            xlabel=Code Rate $R$, 
            ylabel=$\Omega_{\text{opt}}$,  
            grid=both, grid style={gray!10},
            tick align=outside, tickpos=left, 
            legend columns =5,
            legend style={at={(0.495,1.05)},anchor=south}],
        ]
   	\addplot[only marks, Paired-1!70!Paired-2, mark=triangle, semithick, mark size=1.5][error bars/.cd, y dir=both,y explicit] 
        table [x=SNR,y=OPT, y error plus expr=\thisrow{LOW}-\thisrow{OPT},y error minus expr=\thisrow{OPT}-\thisrow{HIGH},] {figures/data2/n64_r_100in1000};

   	\addplot[only marks, Paired-3!70!Paired-4, mark=o, semithick, mark size=1.5][error bars/.cd, y dir=both,y explicit] 
        table [x=SNR,y=OPT, y error plus expr=\thisrow{LOW}-\thisrow{OPT},y error minus expr=\thisrow{OPT}-\thisrow{HIGH},] {figures/data2/n128_r_100in1000};

   	\addplot[only marks, Paired-5!70!Paired-6, mark=square, semithick, mark size=1.5][error bars/.cd, y dir=both,y explicit] 
        table [x=SNR,y=OPT, y error plus expr=\thisrow{LOW}-\thisrow{OPT},y error minus expr=\thisrow{OPT}-\thisrow{HIGH},] {figures/data2/n256_r_100in1000};

   	\addplot[only marks, Paired-7!70!Paired-8, mark=diamond, semithick, mark size=1.5][error bars/.cd, y dir=both,y explicit] 
        table [x=SNR,y=OPT, y error plus expr=\thisrow{LOW}-\thisrow{OPT},y error minus expr=\thisrow{OPT}-\thisrow{HIGH},] {figures/data2/n512_r_100in1000};
        
   	\addplot[only marks, Paired-9!70!Paired-10, mark=pentagon, semithick, mark size=1.5][error bars/.cd, y dir=both,y explicit] 
        table [x=SNR,y=OPT, y error plus expr=\thisrow{LOW}-\thisrow{OPT},y error minus expr=\thisrow{OPT}-\thisrow{HIGH},] {figures/data2/n1024_r_100in1000};
   
             
        \legend{$N=64$,$N=128$,$N=256$,$N=512$,$N=1024$}

    \end{axis}
\end{tikzpicture} 

%% file: figures/approxOmega_demo3.tikz

\begin{tikzpicture}

    \begin{groupplot}[
        group style={group name=group, group size= 3 by 2, horizontal sep=.3cm, vertical sep=2.0cm},
        footnotesize, width=\columnwidth, height=0.55\columnwidth,
        ymode=log,
        ymin = 1.0e-06, ymax=1,
        grid=both, grid style={gray!30},
        tick align=outside, tickpos=left, 
        ]

\nextgroupplot[ylabel={FER}, width = .42\columnwidth, xtick={2, 4, ..., 8},]
\addplot[
    color=Paired-5,
    semithick,
    dashed
]
table{
2.00e+00 5.78000e-01
2.50e+00 4.62000e-01
3.00e+00 3.37900e-01
3.50e+00 2.23100e-01
4.00e+00 1.35100e-01
4.50e+00 7.03000e-02
5.00e+00 3.45000e-02
5.50e+00 1.29000e-02
6.00e+00 3.64033e-03
6.50e+00 1.12590e-03
7.00e+00 3.12848e-04
7.50e+00 4.56244e-05
8.00e+00 6.49683e-06
8.50e+00 1.03694e-06
};
\label{gp:old}

\addplot[
    color=Paired-1,
    semithick
]
table {
2.00e+00 5.96400e-01
2.50e+00 4.81500e-01
3.00e+00 3.59700e-01
3.50e+00 2.41800e-01
4.00e+00 1.50200e-01
4.50e+00 7.92000e-02
5.00e+00 3.85000e-02
5.50e+00 1.51000e-02
6.00e+00 4.59010e-03
6.50e+00 1.22076e-03
7.00e+00 3.14893e-04
7.50e+00 4.62337e-05
8.00e+00 6.58844e-06
8.50e+00 1.03694e-06
};
\label{gp:new}

\coordinate (top) at (rel axis cs:0,0);

\nextgroupplot[yticklabels={,,}, width = .42\columnwidth,  xtick={2, 3, ..., 7},]
\addplot[
    color=Paired-5,
    semithick,
    dashed
]
table{
2.50e+00 7.83900e-01
3.00e+00 5.23700e-01
3.50e+00 2.52800e-01
4.00e+00 8.43000e-02
4.50e+00 1.51000e-02
5.00e+00 2.79649e-03
5.50e+00 2.65132e-04
6.00e+00 2.59460e-05
6.50e+00 3.38448e-06
};
\label{gp:old}

\addplot[
    color=Paired-1,
    semithick
]
table {
2.50e+00 8.26800e-01
3.00e+00 6.00700e-01
3.50e+00 3.33200e-01
4.00e+00 1.33800e-01
4.50e+00 3.63000e-02
5.00e+00 6.20000e-03
5.50e+00 5.82126e-04
6.00e+00 4.65788e-05
6.50e+00 3.49610e-06
};
\label{gp:new}

\coordinate (bot) at (rel axis cs:1,0);

\nextgroupplot[yticklabels={,,}, width = .42\columnwidth,  xtick={1, 2, ..., 4},]
\addplot[
    color=Paired-5,
    semithick,
    dashed
]
table{
1.00e+00 6.74500e-01
1.50e+00 3.41000e-01
2.00e+00 1.00400e-01
2.50e+00 1.70000e-02
3.00e+00 1.85563e-03
3.50e+00 1.54854e-04
4.00e+00 1.24607e-05
4.50e+00 1.07960e-06
};
\label{gp:old}

\addplot[
    color=Paired-1,
    semithick
]
table {
1.00e+00 7.20300e-01
1.50e+00 3.97000e-01
2.00e+00 1.26300e-01
2.50e+00 2.21000e-02
3.00e+00 2.68976e-03
3.50e+00 1.65750e-04
4.00e+00 1.24607e-05
4.50e+00 1.07960e-06
};
\label{gp:new}

\coordinate (bot) at (rel axis cs:2,0);


\nextgroupplot[xlabel={$E_b/N_0$ (dB)},ylabel={FER}, width = .42\columnwidth, xtick={0, 1, ..., 4},yshift=+1.0cm]
\addplot[
    color=Paired-5,
    semithick,
    dashed
]
table{
0.00e+00 7.84300e-01
0.50e+00 5.40800e-01
1.00e+00 2.71900e-01
1.50e+00 9.40000e-02
2.00e+00 2.26000e-02
2.50e+00 3.56405e-03
3.00e+00 3.71466e-04
3.50e+00 2.73446e-05
4.00e+00 1.34583e-06
};
\label{gp:old}

\addplot[
    color=Paired-1,
    semithick
]
table {
0.00e+00 7.92800e-01
5.00e-01 5.61400e-01
1.00e+00 2.95000e-01
1.50e+00 1.02700e-01
2.00e+00 2.56000e-02
2.50e+00 2.96138e-03
3.00e+00 3.08680e-04
3.50e+00 2.65713e-05
4.00e+00 1.50058e-06
};
\label{gp:new}

\coordinate (top) at (rel axis cs:0,1);

\nextgroupplot[xlabel={$E_b/N_0$ (dB)},yticklabels={,,}, width = .42\columnwidth,  xtick={0, 1, ..., 4},yshift=+1.0cm]
\addplot[
    color=Paired-5,
    semithick,
    dashed
]
table{
0.00e+00 7.32100e-01
5.00e-01 4.25200e-01
1.00e+00 1.57900e-01
1.50e+00 3.20000e-02
2.00e+00 3.81913e-03
2.50e+00 3.13009e-04
3.00e+00 1.62746e-05
3.5 8.51685e-07
};
\label{gp:old}

\addplot[
    color=Paired-1,
    semithick
]
table {
0.00e+00 7.40200e-01
5.00e-01 4.32900e-01
1.00e+00 1.63300e-01
1.50e+00 3.28000e-02
2.00e+00 3.81913e-03
2.50e+00 4.37484e-04
3.00e+00 1.93077e-05
3.50e+00 1.16655e-06
};
\label{gp:new}

\coordinate (bot) at (rel axis cs:1,1);

\nextgroupplot[xlabel={$E_b/N_0$ (dB)},yticklabels={,,}, width = .42\columnwidth,  xtick={0, 1, ..., 5},yshift=+1.0cm]
\addplot[
    color=Paired-5,
    semithick,
    dashed
]
table{
1.00e+00 6.89200e-01
1.50e+00 2.50000e-01
2.00e+00 3.42000e-02
2.50e+00 1.76261e-03
3.00e+00 6.93048e-05
3.50e+00 3.57036e-06
};

\addplot[
    color=Paired-1,
    semithick
]
table {
1.00e+00 7.34500e-01
1.50e+00 2.87000e-01
2.00e+00 3.95000e-02
2.50e+00 2.18493e-03
3.00e+00 7.90035e-05
3.50e+00 3.06187e-06
};

\coordinate (bot) at (rel axis cs:2,1);

\end{groupplot}

    \node[above = 0.0cm of group c1r1.north] {\footnotesize $PC(64, 16)$};
    \node[above = 0.0cm of group c2r1.north] {\footnotesize $PC(256, 208)$};
    \node[above = 0.0cm of group c3r1.north] {\footnotesize $PC(512, 256)$};

    \node[above = 0.0cm of group c1r2.north] {\footnotesize $PC(512, 128)$};
    \node[above = 0.0cm of group c2r2.north] {\footnotesize $PC(1024, 192)$};
    \node[above = 0.0cm of group c3r2.north] {\footnotesize $PC(1024, 512)$};

    \path (top|-current bounding box.north) -- 
      coordinate(legendpos)
      (bot|-current bounding box.north);
    \matrix[
        matrix of nodes,
        anchor=south,
        draw,
        inner sep=0.2em,
        draw
      ]at([yshift=+.3ex,xshift=-7.0ex]legendpos)
      {
        \ref{gp:old}& \footnotesize TSCF using $\Omega_{\text{opt}}$ &[5pt]
        \ref{gp:new}& \footnotesize TSCF using $\Omega^{*}$  \\};
\end{tikzpicture}

%% file: figures/Rate_vs_FERtheoretical2.tikz
\begin{tikzpicture}[spy using outlines=
    {circle, magnification=1.7, connect spies}]
  \pgfplotsset{
    label style = {font=\fontsize{9pt}{7.2}\selectfont},
    tick label style = {font=\fontsize{7pt}{7.2}\selectfont}
  }
    \begin{axis}[
            footnotesize, width=\columnwidth, height=0.70\columnwidth,    
            xmin=0, xmax=1, xtick={0,0.125, ..., 1},
            xticklabel style={/pgf/number format/.cd,fixed,precision=3},
            ymin=1.0e-18,  ymax=5,
            ymode=log,
            xlabel=Code Rate $R$, 
            height = 5cm,
            grid=both, grid style={gray!10},
            tick align=outside, tickpos=left, 
            legend columns =3,
            legend style={at={(0.495,1.05)},anchor=south}],
        ]
	   	\addplot[Paired-1, semithick, mark = triangle, only marks, mark size = 3.3]
        table [x=rates,y=FER_T,] {figures/data4/n1k_rate_ebno1.0};
        
        \addplot[Paired-1, semithick, mark = triangle, only marks, mark size = 3.3]
        table [x=rates,y=FER_T,] {figures/data4/n1k_rate_ebno1.5};
        
        \addplot[Paired-1, semithick, mark = triangle, only marks, mark size = 3.3]
        table [x=rates,y=FER_T,] {figures/data4/n1k_rate_ebno2.0};
        
        \addplot[Paired-1, semithick, mark = triangle, only marks, mark size = 3.3]
        table [x=rates,y=FER_T,] {figures/data4/n1k_rate_ebno2.5};
        
        \addplot[Paired-1, semithick, mark = triangle, only marks, mark size = 3.3]
        table [x=rates,y=FER_T,] {figures/data4/n1k_rate_ebno3.0};
        
        \addplot[Paired-1, semithick, mark = triangle, only marks, mark size = 3.3]
        table [x=rates,y=FER_T,] {figures/data4/n1k_rate_ebno3.5};
        
        \addplot[Paired-1, semithick, mark = triangle, only marks, mark size = 3.3]
        table [x=rates,y=FER_T,] {figures/data4/n1k_rate_ebno4.0};
        
        \addplot[Paired-1, semithick, mark = triangle, only marks, mark size = 3.3]
        table [x=rates,y=FER_T,] {figures/data4/n1k_rate_ebno4.5};
        
        \addplot[Paired-1, semithick, mark = triangle, only marks, mark size = 3.3]
        table [x=rates,y=FER_T,] {figures/data4/n1k_rate_ebno5.0};
        

        \addplot[Paired-3, semithick, mark = x, only marks]
        table [x=rates,y=FER_1,] {figures/data4/n1k_rate_ebno1.0};
        
        \addplot[Paired-3, semithick, mark = x, only marks]
        table [x=rates,y=FER_1,] {figures/data4/n1k_rate_ebno1.5};
        
        \addplot[Paired-3, semithick, mark = x, only marks]
        table [x=rates,y=FER_1,] {figures/data4/n1k_rate_ebno2.0};
        
        \addplot[Paired-3, semithick, mark = x, only marks]
        table [x=rates,y=FER_1,] {figures/data4/n1k_rate_ebno2.5};
        
        \addplot[Paired-3, semithick, mark = x, only marks]
        table [x=rates,y=FER_1,] {figures/data4/n1k_rate_ebno3.0};
        
        \addplot[Paired-3, semithick, mark = x, only marks]
        table [x=rates,y=FER_1,] {figures/data4/n1k_rate_ebno3.5};
        
        \addplot[Paired-3, semithick, mark = x, only marks]
        table [x=rates,y=FER_1,] {figures/data4/n1k_rate_ebno4.0};
        
        \addplot[Paired-3, semithick, mark = x, only marks]
        table [x=rates,y=FER_1,] {figures/data4/n1k_rate_ebno4.5};
        
        \addplot[Paired-3, semithick, mark = x, only marks]
        table [x=rates,y=FER_1,] {figures/data4/n1k_rate_ebno5.0};
        

        \addplot[Paired-5, semithick, mark = +, only marks]
        table [x=rates,y=FER_2,] {figures/data4/n1k_rate_ebno1.0};
        
        \addplot[Paired-5, semithick, mark = +, only marks]
        table [x=rates,y=FER_2,] {figures/data4/n1k_rate_ebno1.5};
        
        \addplot[Paired-5, semithick, mark = +, only marks]
        table [x=rates,y=FER_2,] {figures/data4/n1k_rate_ebno2.0};
        
        \addplot[Paired-5, semithick, mark = +, only marks]
        table [x=rates,y=FER_2,] {figures/data4/n1k_rate_ebno2.5};
        
        \addplot[Paired-5, semithick, mark = +, only marks]
        table [x=rates,y=FER_2,] {figures/data4/n1k_rate_ebno3.0};
        
        \addplot[Paired-5, semithick, mark = +, only marks]
        table [x=rates,y=FER_2,] {figures/data4/n1k_rate_ebno3.5};
        
        \addplot[Paired-5, semithick, mark = +, only marks]
        table [x=rates,y=FER_2,] {figures/data4/n1k_rate_ebno4.0};
        
        \addplot[Paired-5, semithick, mark = +, only marks]
        table [x=rates,y=FER_2,] {figures/data4/n1k_rate_ebno4.5};
        
        \addplot[Paired-5, semithick, mark = +, only marks]
        table [x=rates,y=FER_2,] {figures/data4/n1k_rate_ebno5.0};
        

        \addplot[white!50!black, thin, dashed]
        table [x=rates,y=FER_T,] {figures/data4/n1k_rate_ebno1.0};
        
        \addplot[white!50!black, thin, dashed]
        table [x=rates,y=FER_T,] {figures/data4/n1k_rate_ebno1.5};
        
        \addplot[white!50!black, thin, dashed]
        table [x=rates,y=FER_T,] {figures/data4/n1k_rate_ebno2.0};
        
        \addplot[white!50!black, thin, dashed]
        table [x=rates,y=FER_T,] {figures/data4/n1k_rate_ebno2.5};
        
        \addplot[white!50!black, thin, dashed]
        table [x=rates,y=FER_T,] {figures/data4/n1k_rate_ebno3.0};
        
        \addplot[white!50!black, thin, dashed]
        table [x=rates,y=FER_T,] {figures/data4/n1k_rate_ebno3.5};
        
        \addplot[white!50!black, thin, dashed]
        table [x=rates,y=FER_T,] {figures/data4/n1k_rate_ebno4.0};
        
        \addplot[white!50!black, thin, dashed]
        table [x=rates,y=FER_T,] {figures/data4/n1k_rate_ebno4.5};
        
        \addplot[white!50!black, thin, dashed]
        table [x=rates,y=FER_T,] {figures/data4/n1k_rate_ebno5.0};
        

        \legend{$\text{FER}_{\text{SC}}$ using $\mathcal{A}$,,,,,,,,,,$\text{FER}^{*}_{\text{SC}}$ using $\mathcal{C}_{\text{\cite{SCF-WCNC18}}}$,,,,,,,,,,$\text{FER}^{*}_{\text{SC}}$ using $\mathcal{C}_{\text{\cite{SCF-GLOBECOM17}}}$,,,,,,,,,,}

        \coordinate (spypoint)      at (axis cs:0.34,1e-03);
        \coordinate (magnifyglass)  at (axis cs:0.10,1e-03);
        \coordinate (spypoint2)     at (axis cs:0.225,6e-15);
        \coordinate (magnifyglass2) at (axis cs:0.45,1e-12);

        \draw [->, semithick, color=white!30!black] (axis cs:0.05,2.0e-16) -- (axis cs:0.25,1.0e-17);
        \node [color=white!30!black] at (axis cs:0.15,5.0e-18) {\tiny $E_b/N_0$};

    \end{axis}

    \spy [magenta, height=1.25cm, width=1.25cm] on (spypoint)
    in node[fill=white] at (magnifyglass);
    \spy [magenta, height=1.5cm, width=1.5cm] on (spypoint2)
    in node[fill=white] at (magnifyglass2);
\end{tikzpicture} 

%% file: figures/results_perf.tikz

\begin{tikzpicture}[spy using outlines=
	{circle, magnification=2.0, connect spies}]
  \pgfplotsset{
    label style = {font=\fontsize{9pt}{7.2}\selectfont},
    tick label style = {font=\fontsize{7pt}{7.2}\selectfont}
  }

    \begin{groupplot}[
        group style={group name=group, group size= 2 by 1, horizontal sep=1.5cm, vertical sep=2.5cm},
        footnotesize, width=\columnwidth, height=0.78\columnwidth,
        ymode=log,
        xlabel={$E_b/N_0$ (dB)},
        grid=both, grid style={gray!30},
        tick align=outside, tickpos=left, 
        ]

\nextgroupplot[ylabel={BER}, width = .50\columnwidth,]

\addplot[
    color=Paired-9,
    mark=square,
    semithick,
    mark size=2,
]
table {
1.00e+00 3.01955e-01
1.25e+00 2.07596e-01
1.50e+00 1.17703e-01
1.75e+00 5.51227e-02
2.00e+00 2.17066e-02
2.25e+00 6.68712e-03
2.50e+00 1.65511e-03
2.75e+00 4.40833e-04
3.00e+00 8.48313e-05
3.25e+00 1.74275e-05
3.50e+00 1.42135e-06
};
\label{gp:SCF}

\addplot[
    color=Paired-7,
    mark=+,
    semithick,
    mark size=2,
]
table {
1.00e+00 3.02991e-01
1.25e+00 2.03787e-01
1.50e+00 1.15299e-01
1.75e+00 5.29483e-02
2.00e+00 2.06417e-02
2.25e+00 5.89981e-03
2.50e+00 1.72557e-03
2.75e+00 3.85884e-04
3.00e+00 8.99602e-05
3.25e+00 1.69831e-05
3.50e+00 1.24474e-06
};
\label{gp:FSCF}

\addplot[
    color=Paired-11,
    mark=diamond,
    semithick,
    mark size=2,
]
table {
1.00e+00 3.11446e-01
1.25e+00 2.17704e-01
1.50e+00 1.28187e-01
1.75e+00 6.28197e-02
2.00e+00 2.53426e-02
2.25e+00 8.43106e-03
2.50e+00 2.28750e-03
2.75e+00 8.06318e-04
3.00e+00 1.35842e-04
3.25e+00 3.10194e-05
3.50e+00 4.37102e-06
};
\label{gp:Pascal}

\addplot[
    color=Paired-5,
    mark=triangle,
    semithick,
    mark size=2,
]
table {
1.00e+00 2.82652e-01
1.25e+00 1.78041e-01
1.50e+00 9.39081e-02
1.75e+00 3.71352e-02
2.00e+00 1.19335e-02
2.25e+00 2.56875e-03
2.50e+00 4.13546e-04
2.75e+00 1.03506e-04
3.00e+00 1.31354e-05
3.25e+00 2.76092e-06
3.50e+00 4.95485e-07
};
\label{gp:TSCF}

\addplot[
    color=Paired-3,
    mark=x,
    semithick,
    mark size=2,
]
table {
1.00e+00 2.97576e-01
1.25e+00 1.93165e-01
1.50e+00 9.95269e-02
1.75e+00 3.98922e-02
2.00e+00 1.18792e-02
2.25e+00 2.73750e-03
2.50e+00 5.76855e-04
2.75e+00 8.63708e-05
3.00e+00 1.71626e-05
3.25e+00 3.05903e-06
3.50e+00 4.34345e-07
};
\label{gp:FTSCF}

\addplot[
    color=Paired-1,
    semithick,
    mark size=2,
]
table {
1.00e+00 2.23647e-01
1.25e+00 1.30405e-01
1.50e+00 6.03458e-02
1.75e+00 2.21766e-02
2.00e+00 6.55377e-03
2.25e+00 1.51568e-03
2.50e+00 3.38301e-04
2.75e+00 3.97373e-05
3.00e+00 8.06694e-06
3.25e+00 1.42421e-06
3.50e+00 1.98299e-07
};
\label{gp:SCO}

\coordinate (spypoint) at (axis cs:3.2,1.0e-5);
\coordinate (magnifyglass) at (axis cs:1.75,2e-6);

\spy [magenta, height=1.5cm, width=1.5cm] on (spypoint)
   in node[fill=white] at (magnifyglass);

\coordinate (top) at (rel axis cs:0,0);

\nextgroupplot[ylabel={FER}, width = .50\columnwidth]


\addplot[
    color=Paired-9,
    mark=square,
    semithick,
    mark size=2,
]
table {
1.00e+00 7.41400e-01
1.25e+00 5.33600e-01
1.50e+00 3.18300e-01
1.75e+00 1.53000e-01
2.00e+00 6.20000e-02
2.25e+00 2.04000e-02
2.50e+00 5.20000e-03
2.75e+00 1.25748e-03
3.00e+00 2.97338e-04
3.25e+00 6.83533e-05
3.50e+00 7.12159e-06
};
\label{gp:SCF}

\addplot[
    color=Paired-7,
    mark=+,
    semithick,
    mark size=2,
]
table {
1.00e+00 7.41900e-01
1.25e+00 5.25300e-01
1.50e+00 3.12900e-01
1.75e+00 1.48500e-01
2.00e+00 6.05000e-02
2.25e+00 1.84000e-02
2.50e+00 5.20000e-03
2.75e+00 1.20247e-03
3.00e+00 3.10451e-04
3.25e+00 6.60996e-05
3.50e+00 6.84894e-06
};
\label{gp:FSCF}

\addplot[
    color=Paired-11,
    mark=diamond,
    semithick,
    mark size=2,
]
table {
1.00e+00 7.66700e-01
1.25e+00 5.63700e-01
1.50e+00 3.51200e-01
1.75e+00 1.78400e-01
2.00e+00 7.63000e-02
2.25e+00 2.64000e-02
2.50e+00 7.50000e-03
2.75e+00 2.45664e-03
3.00e+00 5.17716e-04
3.25e+00 1.24853e-04
3.50e+00 1.81439e-05
};
\label{gp:Pascal}

\addplot[
    color=Paired-5,
    mark=triangle,
    semithick,
    mark size=2,
]
table {
1.00e+00 6.94900e-01
1.25e+00 4.63800e-01
1.50e+00 2.58100e-01
1.75e+00 1.08600e-01
2.00e+00 3.77000e-02
2.25e+00 8.50000e-03
2.50e+00 1.76261e-03
2.75e+00 4.60723e-04
3.00e+00 9.25475e-05
3.25e+00 2.09449e-05
3.50e+00 4.44774e-06
};
\label{gp:TSCF}

\addplot[
    color=Paired-3,
    mark=x,
    semithick,
    mark size=2,
]
table {
1.00e+00 7.29400e-01
1.25e+00 5.02100e-01
1.50e+00 2.81300e-01
1.75e+00 1.21200e-01
2.00e+00 3.90000e-02
2.25e+00 9.90000e-03
2.50e+00 2.18493e-03
2.75e+00 3.71427e-04
3.00e+00 9.29228e-05
3.25e+00 2.16801e-05
3.50e+00 4.48971e-06
};
\label{gp:FTSCF}

\addplot[
    color=Paired-1,
    semithick,
    mark size=2,
]
table {
1.00e+00 6.59430e-01
1.25e+00 4.27870e-01
1.50e+00 2.20750e-01
1.75e+00 9.03500e-02
2.00e+00 2.95000e-02
2.25e+00 7.42000e-03
2.50e+00 1.73000e-03
2.75e+00 2.99287e-04
3.00e+00 7.08938e-05
3.25e+00 1.43373e-05
3.50e+00 2.92422e-06
};
\label{gp:SCO}

\coordinate (spypoint2) at (axis cs:3.1,1.0e-4);
\coordinate (magnifyglass2) at (axis cs:1.75,2.2e-5);

\spy [magenta, height=1.5cm, width=1.5cm] on (spypoint2)
   in node[fill=white] at (magnifyglass2);

\coordinate (bot) at (rel axis cs:1,0);

\end{groupplot}

    \path (top|-current bounding box.north) -- 
      coordinate(legendpos)
      (bot|-current bounding box.north);
    \matrix[
        matrix of nodes,
        anchor=south,
        draw,
        inner sep=0.2em,
        draw
      ]at([yshift=+.3ex,xshift=-0.0ex]legendpos)
      {
        \ref{gp:SCF}& 	 \footnotesize SCF \cite{SCFlip14} &[5pt]
        \ref{gp:FSCF}& 	 \footnotesize Fast-SCF \cite{FastSCF-TCAS-I} &[5pt]
        \ref{gp:Pascal}& \footnotesize Fast-SSC-Flip \cite{FastSCFlip_WCNC18} \\
        \ref{gp:TSCF}& 	 \footnotesize TSCF \cite{SCFlip_TCOM18} &[5pt]
        \ref{gp:FTSCF}&  \footnotesize Fast-TSCF &[5pt]
        \ref{gp:SCO}& 	 \footnotesize SCO  \\};
\end{tikzpicture}

%% file: figures/results_iter.tikz
\begin{tikzpicture}
    \begin{axis}[
            footnotesize, width=\columnwidth, height=4.75cm,    
            xmin=0.9, xmax=3.6, xtick={0,0.25, ..., 3.5},
            ymode=log,
            xlabel=$E_b/N_0$ (dB), 
            ylabel=Avg. $\#$ of Decoding Steps,  
            grid=both, grid style={gray!10},
            tick align=outside, tickpos=left, 
            legend columns =3,
            legend style={at={(0.495,1.05)},anchor=south}],
        ]
	\addplot[Paired-9,  semithick, mark = square, mark size = 3.0] table [x=ebno,y=SCF,]              {figures/data5/iter};
	\addplot[Paired-7,  semithick, mark = +, mark size = 3.0] table [x=ebno,y=Fast-SCF,]              {figures/data5/iter};
	\addplot[Paired-11, semithick, mark = diamond, mark size = 3.0] table [x=ebno,y=Fast-SSC-Flip,]   {figures/data5/iter};
	\addplot[Paired-5,  semithick, mark = triangle, mark size = 3.0] table [x=ebno,y=TSCF,]           {figures/data5/iter};
	\addplot[Paired-3,  semithick, mark = x, mark size = 3.0] table [x=ebno,y=Fast-TSCF,]             {figures/data5/iter};

        \legend{SCF \cite{SCFlip14}, Fast-SCF \cite{FastSCF-TCAS-I}, Fast-SSC-Flip \cite{FastSCFlip_WCNC18}, TSCF \cite{SCFlip_TCOM18}, Fast-TSCF}

    \end{axis}
\end{tikzpicture} 